\documentclass[11pt]{article}
\usepackage[a4paper,margin=1in]{geometry}
\usepackage[utf8]{inputenc}
\usepackage[T1]{fontenc}
\usepackage{lmodern}
\usepackage{amsmath,amssymb}
\usepackage{graphicx}
\usepackage{xcolor}
\usepackage{authblk}
\usepackage[numbers,sort&compress]{natbib}
\usepackage{caption}
\usepackage{hyperref}
\hypersetup{colorlinks=true,linkcolor=blue,citecolor=blue,urlcolor=blue}

\title{Equilibrium gigahertz acoustics reveals long-range confinement in liquids}
\author[1]{Ievgeniia Chaban}
\author[2]{Thomas Pezeril}
\affil[1]{Laboratoire de M\'ecanique des Solides, UMR CNRS 7649, E\'cole Polytechnique, Institut Polytechnique de Paris, 91128 Palaiseau, France}
\affil[2]{Institut de Physique de Rennes, UMR CNRS 6251, Universit\'e Rennes, 35042 Rennes, France}
\date{}

\begin{document}
\maketitle

\begin{abstract}
Understanding how the mechanical properties of liquids confined within nanometer-scale gaps differ from bulk behavior is central to biophysics, lubrication, catalysis, electrochemistry, and surface science. Yet the characterization of ultrathin confined liquids remains challenging, as many existing approaches rely on destructive or intrusive contact-based techniques, mostly measuring the liquid flow in the low frequency regime. Here, we present a non-invasive, all-optical technique based on ultrafast laser ultrasonics that probes confined liquids at equilibrium in the gigahertz frequency range. The method measures the phase and amplitude of time-domain Brillouin scattering signals transmitted through liquid layers whose thickness is varied step by step with subnanometer effective sampling. Supported by numerical modeling of acoustic propagation and optical detection, these signals allow us to extract the thickness-dependent acoustic velocity and attenuation of confined liquids. We show that nanometric confinement modifies the GHz acoustic response of glycerol, the liquid crystal 8CB, and a butyl-based ionic liquid over unexpectedly long spatial scales. These effects extend from a few nanometers to several tens of nanometers and reveal bound interfacial layers, acoustic stiffening, and enhanced solid-like behavior under confinement. Our results open a route to probing liquid confinement in a scarcely explored regime: dynamically measured at gigahertz frequencies, yet sufficiently weakly perturbative to preserve the equilibrium confined state.
\end{abstract}

\noindent\textbf{Keywords:} Liquid nanoconfinement ; laser ultrasonics ; interfacial mechanics

\section*{Significance Statement}
Mechanical properties of liquids confined between solid surfaces are often altered from the properties of bulk liquids. Liquids under nanoconfinement undergo liquid-to-solid transitions resulting in molecules solid-like organization. In our study, we reveal experimental evidence, obtained using an all-optical laser ultrasonics technique, that various liquids exhibit GHz solid-like behavior under surface nanoconfinement. Unattainable by conventional methods, these results highlight an unexpected long range confinement, ranging from a few to several tens of nanometers, that can be explained by the in-equilibrium measurements provided by the ultrasonic technique.

\section*{Introduction}
Liquids confined within nanometric gaps are ubiquitous in natural and technological systems, from biological interfaces and membranes to lubrication, nanofluidics, electrochemical devices, and porous materials. In such geometries, even simple liquids can differ markedly from their bulk behavior \cite{Israelachvili_1988,Klein_1995,Demirel_1996}. When squeezed between solid surfaces down to few molecular layers, liquids may develop interfacial ordering \cite{Hulsman_1997}, density oscillations \cite{Steinruck_2014}, and molecular layering \cite{Mezger_2008}, leading to modified transport, enhanced mechanical stiffness, and, in some cases, an apparent solid-like response over distances ranging from a few molecular layers to several tens of nanometers \cite{Fukuchi_2001,Nanoindentation_2006,Goyon_2008,Sharma_2008,Khan_2010,Mante_2014,Gebbie_2017,Comtet_2017,Lhermerout_2018,Garcia_2018,Weiss_2019,Laine_2020,Schlaich_2021,Scalfi_2021,Holey_2024,Mante_2024,Kurihara_2025}. Understanding how confinement modifies the mechanical response of liquids is, therefore, essential for describing interfacial friction, adhesion, dissipation, and energy transport at small scales.

However, probing this response remains challenging. The relevant volumes are extremely small, and the measured properties are governed by a subtle interplay between molecular structure, surface chemistry, charge, roughness, and wetting. A variety of experimental approaches, including scanning probe microscopy \cite{Ducker1991, Butt1991}, surface forces apparatus measurements \cite{Klein_1995,Heuberger_2001,Israelachvili2010,Charlaix2012,Garcia_2018,Perkin2018,Laine_2020,Perkin2023}, diffraction techniques \cite{Reichert2017, Sentker2018, Perakis2018}, nonlinear optical spectroscopy \cite{Loughnane2000, Farrer2003}, and infrared or terahertz spectroscopy \cite{Huber2014}, have provided important insights into interfacial structure and slow mechanical relaxation. Yet many of these methods rely on static or quasi-static loading, require direct mechanical contact, or infer elastic properties indirectly from structural or spectroscopic observables. As a result, the high-frequency mechanical response of molecularly confined liquids remains largely unexplored, although this dynamical regime is expected to reveal how liquids respond when molecular rearrangements cannot fully relax during the measurement.

Laser ultrasonics provides a route to this dynamic regime. In this approach, an ultrashort optical pulse absorbed in a thin transducer layer produces a rapid thermoelastic stress that launches a coherent acoustic pulse \cite{Thompsen1984}. The pulse propagates through the sample and interacts with buried interfaces or nanoscale layers, while a delayed optical probe detects the resulting transient displacement or reflectivity modulation. Because both excitation and detection are optical, the method is contact-free and compatible with confined or nanostructured geometries. Its picosecond timescale gives access to GHz-frequency mechanical properties, allowing liquids to be probed under conditions in which molecular rearrangements may not fully relax. Previous laser-ultrasonic studies have shown that laser-generated acoustic waves can interrogate liquids in submicrometer gaps \cite{Pezeril2009,Klieber2012}. More recently, reflection-mode acoustic measurements across solid-liquid interfaces have revealed liquid structuring through changes in acoustic energy transfer \cite{Chi-Kuang_2020,Uthe_2022}. However, applications to molecularly confined liquids remain limited. In particular, it remains unclear how the acoustic response evolves when the liquid thickness is tuned in a controlled manner down to the nanometric regime, where layering, surface-induced ordering, and solid-like mechanical behavior are expected to emerge.

\begin{figure*}[ht!]
\centering
\includegraphics[width=0.7\textwidth]{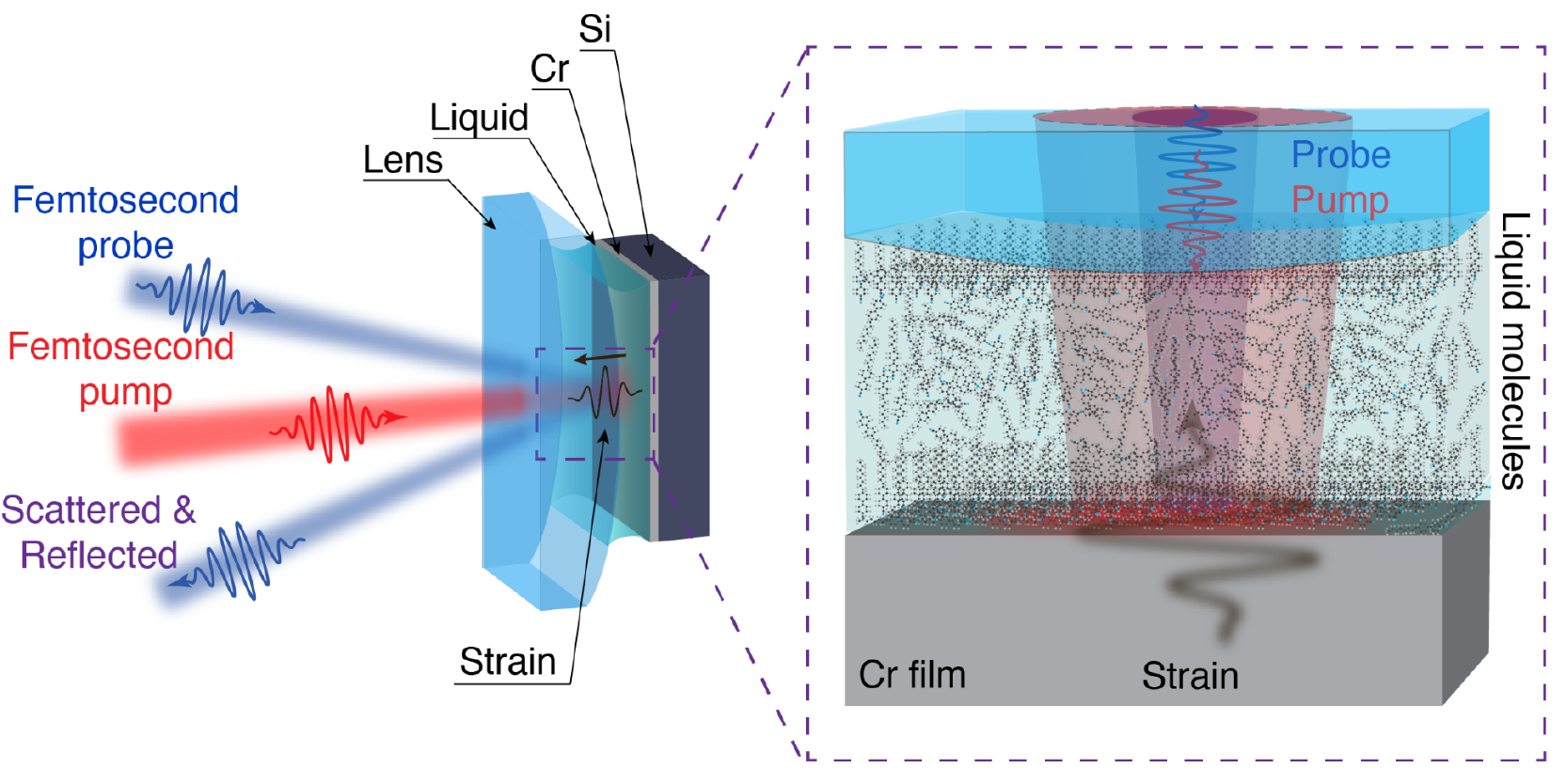}
\caption{Schematic representation of the nanoconfinement measurements. A liquid layer is confined between a flat silicon substrate coated with an optoacoustic transducer chromium film and a plano-convex glass lens with a large radius of curvature. This flat-curved geometry allows the liquid to flow within the confined region while providing a well-controlled nanometric gap. The liquid thickness is fine tuned with subnanometer resolution using a piezoelectric actuator onto which the silicon substrate is glued. A femtosecond pump pulse is partially absorbed in the chromium film, where it launches GHz ultrasonic pulses. After transmission across the confined liquid layer, the acoustic pulses propagate in the glass lens, where they are detected by a delayed femtosecond probe pulse.}
\label{figure1}
\end{figure*}

Here, we use laser ultrasonics to investigate the mechanical response of several liquids confined at nanometric length scales. A thin chromium film deposited on a silicon substrate serves as an optoacoustic transducer. Upon optical excitation, it launches a longitudinal acoustic pulse that propagates through the structure and probes the confined liquid layer. By analyzing the time-of-flight, amplitude, and phase of the detected acoustic signal, we access the effective acoustic response of the confined liquid in a non-invasive manner. We report GHz signatures of confinement in several liquids, including glycerol, 4-N-octyl-cyanobiphenyl (C$_{21}$H$_{26}$N), a liquid crystal known under the name 8CB, and a butyl-based ionic liquid, under weakly perturbative conditions that preserve the equilibrium confined state. Remarkably, these confinement effects extend over long-range spatial scales, from several nanometers to tens of nanometers, while being probed in the GHz-frequency regime. This methodology opens a window onto liquid confinement in a scarcely explored regime: dynamically measured at ultrahigh frequency, yet sufficiently gentle to probe the undisrupted equilibrium state of the confined liquid.

\section*{Results}

In this work, we focus on the mechanical characterization of confined liquid layers using an all-optical, non-contact, non-invasive technique to laser-excite and laser-detect ultrasounds in the GHz frequency range. The laser-excitation mechanism relies on the thermoelastic generation of acoustic waves in a metallic transducer. In our geometry, a liquid layer is confined between a flat silicon substrate coated with a chromium metallic film and a plano-convex glass lens (Fig. \ref{figure1}). A fraction of a femtosecond laser pulse, namely the pump pulse, is absorbed within the optical skin-depth of the chromium optoacoustic transducer. The resulting rapid temperature rise produces an impulsive lattice expansion, which launches a picosecond strain pulse into the surrounding media. Because of its short duration, this acoustic pulse contains frequency components in the tens of gigahertz range. As in any multilayer acoustic stack, the strain pulse is partially reflected and transmitted at each interface. The strain pulse generated in the chromium film is partially transmitted across the chromium/liquid interface, reverberates within the confined liquid layer, and gets partially transmitted into the glass lens. Once in the transparent glass, the propagating strain pulse is optically detected  by a delayed femtosecond laser pulse, namely the probe pulse. 

The key feature of the optical detection is the interference between probe light scattered by the propagating acoustic pulse through the photoelastic effect and light reflected from the sample interfaces. This mechanism, known as time-domain Brillouin scattering (TDBS), produces sinusoidal oscillations in the detected probe intensity. The oscillation frequency, referred to as the Brillouin frequency, is set by the acoustic velocity, the refractive index of the glass, the probe wavelength, and the optical scattering geometry.

\section*{Case study: glass forming liquids, glycerol}

\begin{figure*}[ht!]
\centering
\includegraphics[width=0.8\textwidth]{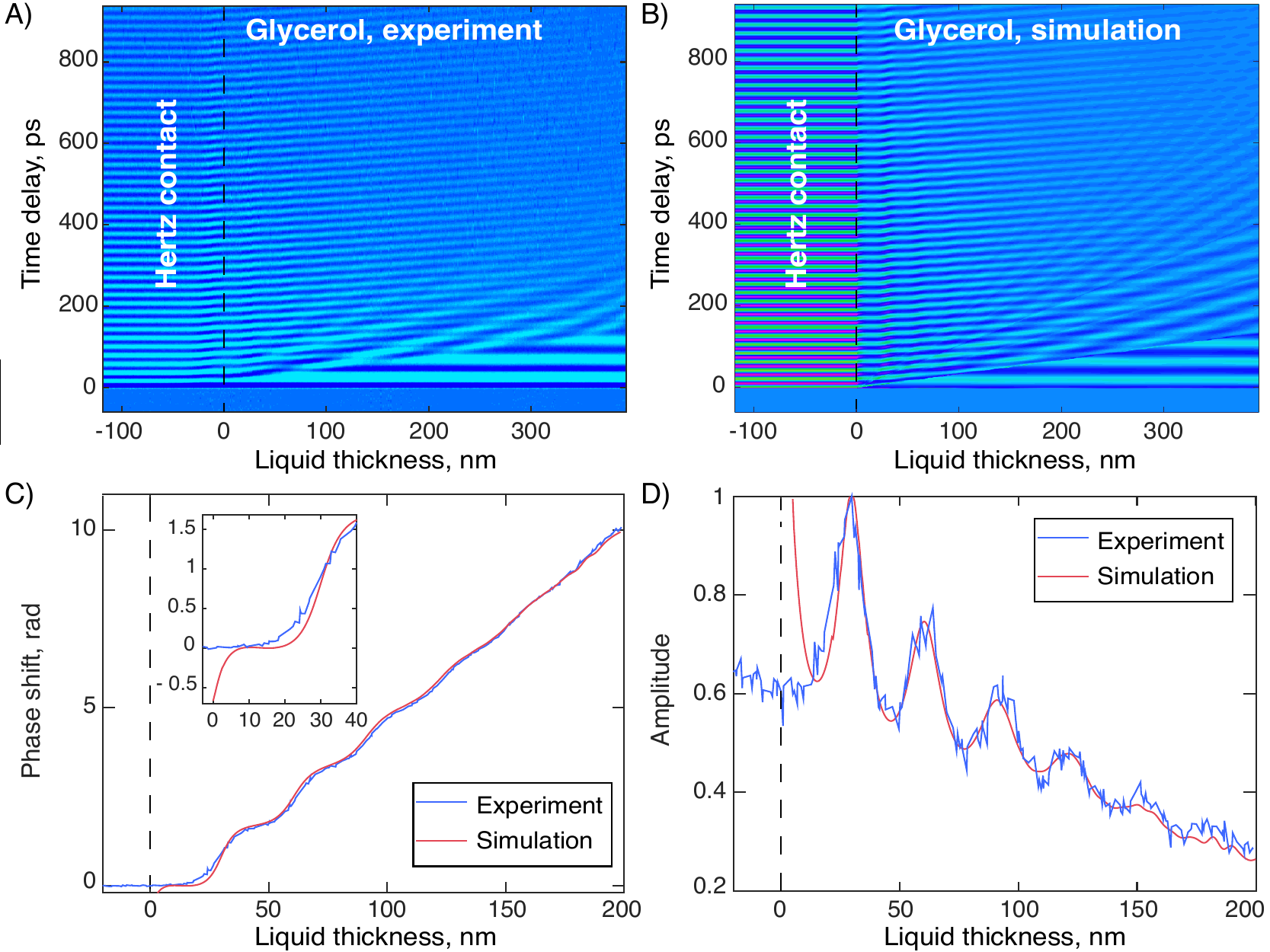}
\caption{Time-domain Brillouin scattering measurements and numerical simulations for confined glycerol. (A) Interpolated two-dimensional map of the measured TDBS signals as a function of pump--probe delay and liquid thickness. Apparent negative thicknesses correspond to the contact regime, in which the chromium film is directly loaded by the glass lens under weak, reversible Hertz contact conditions. The high-frequency oscillations arise from Brillouin scattering of the acoustic pulse propagating in the glass lens. (B) Corresponding numerical simulation of acoustic propagation in the multilayer structure and optical detection in the glass lens. (C) Thickness-dependent phase shift of the glass Brillouin oscillations, from which the acoustic time delay across the confined liquid layer is extracted. (D) Thickness-dependent Brillouin amplitude, which encodes acoustic transmission and damping in the confined liquid.}
\label{figure2}
\end{figure*}

An example of such a time-domain Brillouin scattering signal is shown in Fig.~\ref{figure2}A. The figure displays a 2D map of the probe reflected intensity as a function of pump-probe delay and liquid thickness. The pronounced oscillations are observed over a broad range of thicknesses. By convention, negative thicknesses correspond to the regime in which the chromium film is brought into direct contact with the glass lens, governed by Hertzian contact mechanics. In this configuration, the curved lens applies a local mechanical load to the substrate, producing a reversible elastic bending of the contact region. This regime is therefore only loosely related to nanoindentation: unlike conventional indentation experiments, no sharp tip is used and the deformation remains gentle, purely elastic and well described by Hertzian contact mechanics. In this framework, the elastic interaction between the chromium film and the glass lens is modeled as the frictionless contact of two smooth, curved elastic bodies, leading to a finite circular contact area that grows nonlinearly with load and a characteristic semi-elliptical pressure distribution.

In this contact regime, the ripples in Fig.~\ref{figure2}A are bright and nearly horizontal, with a frequency $\nu_1$ of 47.4~GHz, corresponding to the Brillouin frequency of the glass lens. As the liquid thickness increases, these high-frequency oscillations shift diagonally in the time-thickness map. This shift reflects the acoustic time of flight across the confined liquid layer and is therefore governed by the longitudinal sound velocity in the liquid. At the same time, the oscillation amplitude decreases progressively and eventually vanishes as a result of acoustic attenuation in the liquid at GHz frequencies.

For liquid thicknesses larger than approximately 100~nm, an additional contribution becomes visible, particularly in the 0--100~ps delay range. In this regime, the 47.4~GHz oscillations become weaker and are superimposed with slower, horizontal oscillations, with a frequency $\nu_2$ around 21.1~GHz. These lower-frequency oscillations arise from Brillouin scattering within the liquid itself and can be used, when needed, to determine the sound velocity of the liquid at this frequency. However, the most relevant information on nanoconfinement is contained in the high-frequency 47.4~GHz Brillouin signal transmitted into the glass lens. Accordingly, Fourier analysis of the time-domain data provides access to the thickness-dependent acoustic response of the confined liquid, with the Brillouin phase encoding the acoustic time delay and the Brillouin amplitude reflecting acoustic transmission and damping. However, a direct interpretation of these quantities is complicated by optical and acoustic resonances within the liquid layer, which acts simultaneously as an optical and acoustic cavity. We therefore implemented a numerical model that accounts for acoustic propagation and optical detection in the glass lens through the multilayer structure. This analysis was first applied to glycerol, as shown in Fig.~\ref{figure2}B, which reproduces very well the experimental data in Fig.~\ref{figure2}A.

Figures~\ref{figure2}C and \ref{figure2}D summarize the Fourier analysis of the Brillouin signal detected in the glass lens. For each liquid thickness, we computed the fast Fourier transform of the time-domain traces shown in Fig.~\ref{figure2}A that yields two thickness-dependent quantities: the Brillouin phase and the Brillouin amplitude. The Brillouin phase in Fig.~\ref{figure2}C displays an approximately linear increase with liquid thickness, superimposed with periodic modulations at small thicknesses. In the simplified limit where acoustic reverberations in the liquid layer are neglected, the phase shift can be written as \cite{Klieber2012}
\begin{equation}
\phi = 2\pi (\nu_1 - \nu_2) \frac{d}{c_1},
\label{phase}
\end{equation}
where $d$ is the liquid thickness, $c_1$ is the longitudinal acoustic velocity of the liquid at the glass Brillouin frequency $\nu_1$, and $\nu_2$ is the Brillouin frequency associated with scattering in the liquid itself. This expression shows that the phase slope is inversely proportional to the liquid speed of sound. Importantly, because the liquid response is viscoelastic and therefore frequency dependent, $c_1$ corresponds to the sound velocity at $\nu_1$, not at the native liquid Brillouin frequency $\nu_2$. The two frequencies can differ substantially, and so can the corresponding acoustic velocities. In Eq.~\ref{phase}, $\nu_1$ accounts for the acoustic time delay imposed by propagation through the liquid layer, whereas $\nu_2$ accounts for the optical phase contribution associated with Brillouin scattering in the liquid. Assuming that the sound velocity of glycerol is independent of thickness in the bulk-like regime above approximately 50~nm, the measured phase slope of $0.057$~rad$\cdot$nm$^{-1}$ gives $c_1 = 2900$~m$\cdot$s$^{-1}$, in good agreement with previous measurements \cite{Pezeril2009,Klieber2012}. To describe the phase modulation observed at smaller thicknesses, which arises from acoustic reverberations within the confined liquid cavity, we implemented a numerical model using a constant glycerol sound velocity. As shown in Fig.~\ref{figure2}C, the model reproduces both the linear phase slope at larger thickness and the oscillatory deviations observed at smaller thickness. These oscillations originate from the multiple acoustic reflections occurring inside the liquid layer. After crossing the liquid cavity, the initially generated strain pulse emerges into the glass as a train of pulses with decreasing amplitudes, separated by the acoustic round-trip time in the liquid layer (see Materials and Methods). When this time separation becomes commensurate with the Brillouin period detected in the glass, the transmitted signal exhibits an acousto-optic resonance. This effect is clearly visible in the Brillouin amplitude shown in Fig.~\ref{figure2}D, where the amplitude follows a damped oscillatory dependence on liquid thickness. The damping of these oscillations with increasing thickness reflects acoustic attenuation in the liquid. The thickness period of the amplitude modulation is approximately $T_d = 28.8$~nm, and it should match  half of the acoustic wavelength in the liquid at $\nu_1$, i.e., $T_d =\lambda_1/2 = c_1/(2\nu_1)$. This relation provides an independent, although less accurate, estimate of the sound velocity. Using $c_1 = 2\nu_1T_d$ gives $c_1 = 2730$~m$\cdot$s$^{-1}$. From the numerical simulations of both the amplitude and phase, we better fit the speed of sound of 2920~m$\cdot$s$^{-1}$, which is more reliable than that inferred from the amplitude period alone, and extract the attenuation coefficient $\alpha \simeq 7 \times 10^6$~m$^{-1}$ at the glass Brillouin frequency.

A clear deviation from the constant-velocity model appears at the onset of the first resonance region, between approximately 20 and 40~nm. In this range, the measured amplitude modulation shifts relative to the numerical simulations prediction, indicating that the confined liquid no longer behaves as a bulk-like acoustic cavity. A corresponding discrepancy is also visible in the Brillouin phase inset of Fig.~\ref{figure2}C. Together, these observations reveal a measurable departure from bulk glycerol properties at tens-of-nanometers confinement. From our estimates, the effective acoustic response differs from the bulk value by approximately 10\% in this thickness range. The deviation becomes even more pronounced below approximately 5~nm. In this ultrathin regime, the liquid no longer displays the oscillatory response expected for a freely flowing, bulk-like liquid cavity. Instead, the Brillouin amplitude evolves as if the liquid acted as an acoustic matching layer mechanically coupled to the chromium film (Fig.~\ref{figure2}D). The data are best reproduced by a matching bound interfacial layer that remains coupled to the substrate at 47.4~GHz. This interpretation is consistent with previous laser ultrasonics measurements in acoustic reflection geometry \cite{Chi-Kuang_2020,Uthe_2022}. Under nanometric confinement, this high-frequency elastic response could be amplified by interfacial coupling, leading to the formation of a solid-like confined layer. The same matching behavior is observed from the contact regime up to thicknesses of order 10--15~nm in Figures~\ref{figure2}C and \ref{figure2}D, where the Brillouin amplitude varies monotonically rather than sinusoidally. These observations indicate that, at 47.4~GHz, the confined liquid no longer behaves as a simple acoustic cavity but as a mechanically bound interfacial medium.

Several points support the interpretation that the observed deviations reflect confinement-induced changes in the GHz mechanical response of the liquid. The measurements are performed at fixed thickness under equilibrium conditions, rather than during driven flow or squeeze-out dynamics, so the response probes the confined state itself, which is very different than previous studies in the field of liquid confinement \cite{Bresme2022_review}. However, photoinduced effects must be considered. Femtosecond pump and probe pulses, even at low fluence, could in principle induce molecular alignment, interfacial rearrangement, or photochemical modification near the chromium surface, particularly because the probe wavelength lies close to the UV and the metallic layer may catalyze interfacial reactions. However, such effects are unlikely to dominate the present observations: the signals are reversible, no cumulative evolution is observed during repeated measurements, and the estimated temperature rise remains below 10~K \cite{ChabanRSI}. A purely thermal origin would, moreover, tend to soften the liquid rather than produce the enhanced acoustic coupling observed at small gaps.

The most natural interpretation is therefore that confinement slows molecular relaxation and enhances the effective viscoelastic response at 47.4~GHz. At this frequency, molecular rearrangements may be too slow to fully relax during one acoustic cycle, causing the confined liquid to respond more elastically than in the low-frequency limit. This effect can be amplified near solid boundaries, where adsorption, surface-induced ordering, reduced mobility, and nanoscale roughness may promote the formation of a mechanically bound interfacial layer. In our geometry, the root-mean-square (RMS) roughness is comparable on both confining surfaces, with values of 0.98~nm for the chromium film and 1.04~nm for the glass lens. These values are small compared with the long-range confinement scale. In this regime, the confined layer behaves as an acoustic matching layer rather than as a freely flowing liquid cavity. Surface forces and structural disjoining pressures may further modify the density, compressibility, and viscosity over nanometric to tens-of-nanometers distances. Together, the phase and amplitude deviations point to a genuine confinement-induced modification of the liquid mechanics, revealed here in a high-frequency yet weakly perturbative regime.

\section*{Long-range confinement in ionic liquids and liquid crystals}

\begin{figure*}[ht!]
\centering
\includegraphics[width=0.8\textwidth]{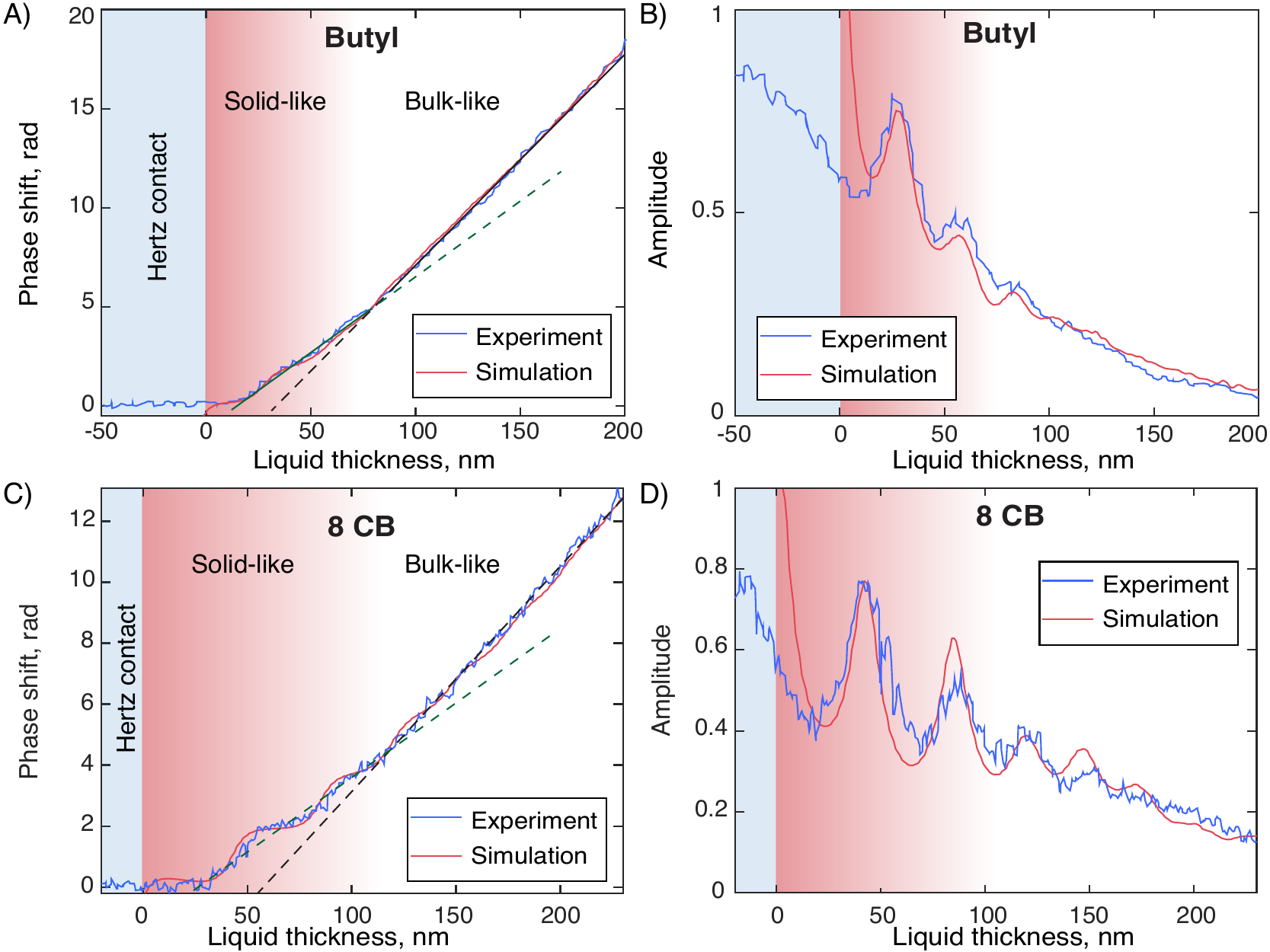}
\caption{Time-domain Brillouin scattering measurements and numerical simulations for the butyl-based ionic liquid and liquid crystal 8CB. (A) Thickness-dependent Brillouin phase shift and (B) Brillouin amplitude for the $Butyl$. The data reveal two acoustic regimes separated by a confinement-induced crossover. (C) Thickness-dependent Brillouin phase shift and (D) Brillouin amplitude for 8CB, showing a similar transition between bulk-like and confined-state responses.}
\label{figure3}
\end{figure*}

We performed similar measurements on a butyl-based ionic liquid and on the liquid crystal 8CB (4-cyano-4'-octylbiphenyl). The corresponding data are shown in Fig.~\ref{figure3}. For the ionic liquid (Fig.~\ref{figure3}A), the Brillouin phase displays two nearly linear regimes with distinct slopes, separated by a crossover around 80~nm. We assign the large-thickness regime to the bulk-like amorphous liquid response, whereas the smaller-thickness regime indicates a confinement-modified state with a larger effective acoustic velocity. The Brillouin amplitude shown in Fig.~\ref{figure3}B does not exhibit a similarly sharp two-regime behavior, suggesting that the acoustic attenuation changes less strongly than the sound velocity across this confinement transition. Numerical simulations reproduce the experimental phase and amplitude responses when the acoustic velocity is described by a smooth thickness-dependent profile. The best agreement is obtained using an error-function-like transition from a high confined-state velocity of approximately 2850~m~s$^{-1}$ below 50~nm to a bulk-like velocity of approximately 1940~m~s$^{-1}$ over a crossover width of about 30~nm and a constant acoustic attenuation coefficient $\alpha \simeq 12 \times 10^6$~m$^{-1}$. These values indicate a pronounced stiffening of the ionic liquid under long-range confinement. Similar long-range structural and mechanical changes have been reported for ionic liquids using surface force apparatus measurements \cite{Garcia_2018}, suggesting that the effect is not solely a consequence of the GHz probing frequency. As observed for glycerol, the response below approximately 10~nm cannot be fully captured by a simple liquid-cavity model with ideal boundaries. Instead, the data point to the presence of strongly bound molecules at the confining surfaces, forming an interfacial layer that acts as an efficient acoustic matching medium.

The measurements on 8CB (Fig.~\ref{figure3}C and \ref{figure3}D) show a similarly clear separation between two structural regimes. The best agreement with the data is obtained using an error-function-like velocity profile, with a high confined-state velocity of approximately 4100~m~s$^{-1}$ below 70~nm and a bulk-like velocity of approximately 2600~m~s$^{-1}$ at larger thicknesses, connected by a broad crossover of about 50~nm. In contrast to the ionic liquid, the acoustic attenuation also changes substantially, matching the error-function-like, ranging from $\alpha \simeq 0.7 \times 10^6$~m$^{-1}$ to $5 \times 10^6$~m$^{-1}$. The simultaneous increase in acoustic velocity and strong variation in attenuation indicate a pronounced confinement-induced modification of both the elastic and dissipative response of 8CB. We previously measured the acoustic velocity and attenuation of 8CB in the thick-film regime using TDBS combined with lateral scanning \cite{Chaban2020}. The present piezo-controlled measurements agree closely with those independent values, validating the bulk-like limit of the analysis at large thickness. The already high velocity of approximately 2600~m~s$^{-1}$ is consistent with the intrinsic molecular order of 8CB in its liquid-crystalline state. Below the confinement threshold of 70~nm, however, the further increase to approximately 4100~m~s$^{-1}$ reveals a distinct confined regime characterized by enhanced molecular organization and a more solid-like GHz mechanical response.

\section*{Concluding Remarks}

In summary, we have shown that TDBS, combined with numerical simulations, provides a sensitive all-optical probe of the GHz mechanical response of liquids under nanometric confinement. Across glycerol, a butyl-based ionic liquid, and the liquid crystal 8CB, we observe confinement-induced deviations from bulk behavior over length scales ranging from a few nanometers to several tens of nanometers. These deviations reveal distinct confined regimes, from interfacial bound layers in the thinnest gaps to long-range solid-like or highly ordered states at larger separations. By accessing the acoustic phase and amplitude without imposing flow or static mechanical loading, this approach opens a route to measuring confined liquids in a high-frequency yet weakly perturbative equilibrium regime. More broadly, our results establish TDBS as a powerful method for probing the mechanical consequences of liquid confinement in regimes that remain difficult to access with conventional techniques.

\section*{Materials and Methods}

\subsection*{TDBS experiments}
TDBS experiments were performed using an ultrafast optical pump-probe setup. We used a femtosecond Ti:sapphire regenerative amplifier (Coherent RegA 9000) delivering $\sim$260~fs pulses at 260~kHz repetition rate, centered at 790~nm. The output beam was split into a pump and a probe beams. The pump was modulated by an acousto-optic modulator at a subharmonic frequency of the laser repetition rate. After the acousto-optic modulator, the pump beam passed through a motorized delay stage, enabling continuous modification of the time difference between the pump and probe pulses of different optical paths. At the sample, the pump beam was focused onto the chromium film with a Gaussian spatial beam profile having a full width at half maximum (FWHM) of 100~$\mu$m. A low pump fluence of 0.5~mJ/cm$^{2}$ was used to avoid cumulative heating of the liquid. The probe beam, of much lower intensity, was frequency doubled to 395~nm wavelength via second harmonic generation. The circularly polarized probe was tightly focused at normal incidence onto the sample surface with a spot size of 8~$\mu$m FWHM and spatially overlapped with the pump spot. The spatial overlapwas optimized by maximizing the amplitude of the TDBS signal measured at zero pump--probe delay, $t = 0$~ps. To improve the signal-to-noise ratio, ten consecutive delay scans were acquired for each liquid thickness over a time window from $-100$ to 900~ps. These scans were averaged point by point to obtain a single representative Brillouin signal for each thickness, with a total acquisition time of approximately 2~minutes.

\subsection*{Sample-cell structure}
To measure the mechanical properties of liquids over a wide range of thicknesses using TDBS, we designed a dedicated sample cell that combines a motorized linear stage (Newport, MFA - CC model) with a piezoelectric nanopositioning stage (Piezo-concept, HS-1.10, with 0.01~nm thickness resolution, 10~$\mu$m full range of displacement and 5.2 N/$\mu$m stiffness). The liquid is confined between two solid surfaces, a 40~nm - thick Chromium film deposited on a flat Silicon substrate and a N-BK7 plano-convex lens with a radius of curvature (ROC) of 386~mm. Given the angstrom-level resolution required, surface quality (roughness, impurities and absence of defects) is critical because it can affect the molecular distribution and organization of confined liquid layer. To mitigate these effects, both surfaces were rigorously cleaned with acetone and optical tissue prior to each experiment, and their surface topography were characterized by atomic force microscopy (AFM). A fresh pair of substrates and lenses was used for each experimental campaign. The root-mean-square (RMS) roughness, R$_q$, averaged over the scanned areas, was 0.986~nm for the Chromium/Silicon assembly and 1.04~nm for the lens. AFM scans were performed over 10 $\times$10~$\mu$m$^2$ areas at three different locations on each substrate. Another key geometrical parameter is the ROC of the detection lens. If the ROC is too large, the detection and generation surfaces cannot be brought into direct contact without applying excessive pressure, which risks fracturing the lens; in this regime, the liquid cannot be fully squeezed out to effectively zero thickness. Conversely, if the ROC is too small, the variation of liquid thickness across the probe spot  becomes significant, leading to non-negligible thickness inhomogeneity along the probe beam. For our probe beam of 8~$\mu$m and a ROC of 386~mm, the thickness variation across the probe spot is $\Delta d \approx$(8~$\mu$m)$^2$/(2$\times$0.386~m)=0.083~nm. This value is well below the typical molecular size, indicating that thickness inhomogeneity across the probe spot is negligible.

\begin{figure}
\centering
\includegraphics[width=0.5\columnwidth]{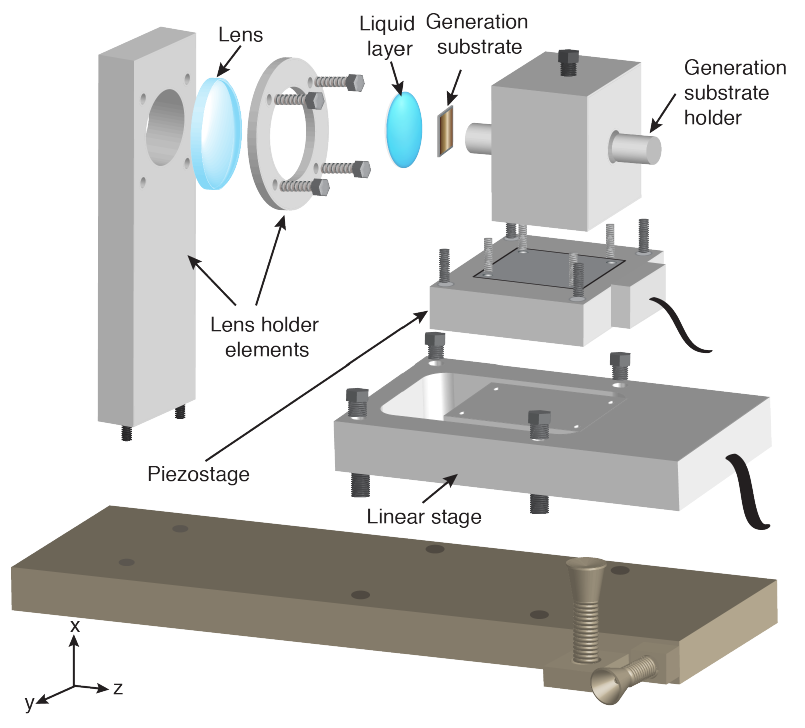}
\caption{Schematic representation of the experimental sample cell. The liquid is confined between a flat chromium-coated silicon substrate and a plano-convex glass lens. The generation substrate is mounted on a motorized linear stage combined with a piezoelectric positioning stage, allowing coarse displacement along the $z$ axis with 0.1~$\mu$m accuracy and ultra-fine displacement with subnanometric resolution to control the liquid thickness. The whole cell is mounted on a three-axis translation stage, enabling lateral positioning and measurements along the $x$ and $y$ axis.}
\label{sfigure1}
\end{figure}

\subsection*{Sample-cell assembly} The assembly of the confined-liquid sample cell requires precise mechanical alignment in addition to careful surface preparation. A simplified schematic of the assembly is shown in Fig.~\ref{sfigure1}. The motorized linear stage and the lens holder were first mounted on a three-axis motorized positioning stage (Newport, M-562-xyz Ultralign). Both components are aligned to be parallel to each other and to the reference axes of the positioning stage. The detection lens is then placed in its dedicated fixed mount and secured by a retaining ring. The ring is fastened to the lens holder using four screws, with the tightening torque distributed as evenly as possible among them. Once this step is completed, the lens holder remains fixed for the entire set of measurements. Under a fume hood, the silicon substrate holding the Chromium layer is glued with epoxy glue from its unpolished back side on top of an aluminum cylindrical holder approximately 8~mm in diameter and 3~cm in length. After waiting a couple of hours that the glue dries and solidifies properly, the cylindrical holder was tightly screwed onto the top side of the aluminum block. Still under a fume hood, a drop of filtered liquid (0.05~$\mu$m pore size filter) was then injected with a needle between the chromium-coated silicon substrate and the glass lens. The cylindrical holder was fixed in such a way that the liquid is gently confined: it cannot flow out, but the two solid surfaces are not yet in direct mechanical contact. At this stage, the liquid film is typically tens of micrometers thick. After assembly and prior to the first TDBS measurement, the entire sample cell is enclosed within a transparent protective box fabricated from 1~cm-thick Plexiglas sheets. This enclosure shields the confined liquid layer from external perturbations, in particular air currents and acoustic vibrations, thereby improving the stability and reproducibility of the TDBS measurements over the duration of the multi-loop TDBS experimental sequences. To reach nominal liquid thicknesses of several hundred nanometers, the two confining surfaces were first brought into near contact using the motorized linear stage. This position was visualized by the appearance of Newton-ring interference fringes, which arise when the liquid thickness becomes comparable to the optical coherence length. Lateral TDBS scans were then performed along the $x$ and $y$ directions across the Newton-ring pattern, following the approach described in \cite{Chaban2020}, to locate the center of the contact region and align it with the pump--probe measurement area. From this reference position, the liquid thickness was gradually increased by retracting the piezo-driven stage, thereby increasing the distance between the lens and the chromium-coated generation substrate. The displacement was applied in small increments, typically 10~nm, to minimize the risk of cavitation. TDBS measurements were performed in real-time in order to track the appearance of Brillouin scattering in the liquid itself, which indicates a liquid thickness in the hundreds of nanometers range (Fig.\ref{figure1}A), which is the starting point of the multi-loop TDBS sequence.

\begin{figure}
\centering
\includegraphics[width=0.5\columnwidth]{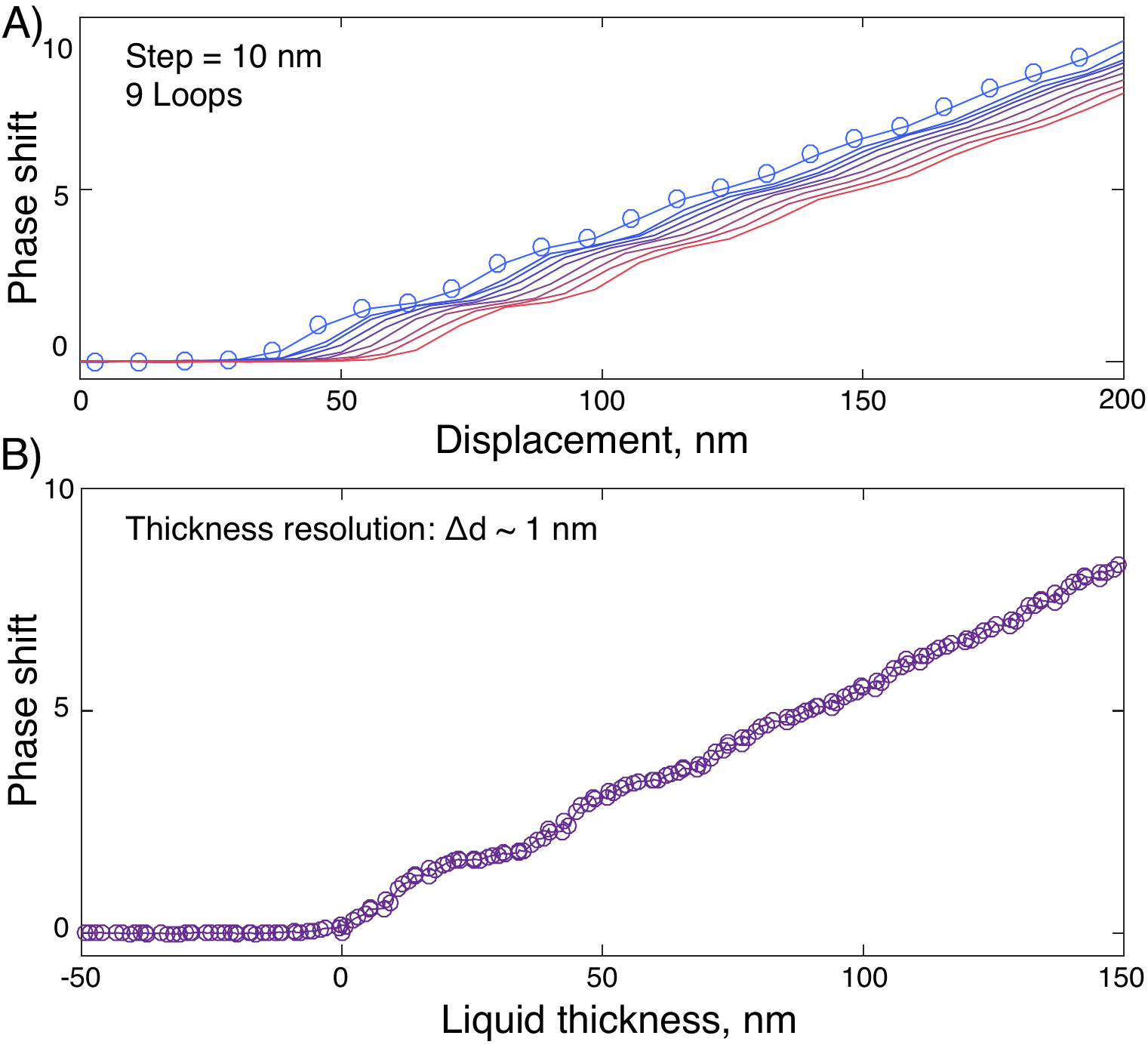}
\caption{Multi-loop extraction of the Brillouin phase at 47.4~GHz as a function of glycerol thickness. (A) Phase data obtained from the nine acquisition loops, each recorded with a nominal 10~nm piezo step. Individual loops are shown separately. (B) Composite thickness dependence obtained by merging all data points from the nine loops. The continuity of the phase evolution across loops confirms the reproducibility of the acquisition and yields an effective thickness sampling of approximately 1~nm.}
\label{sfigure2}
\end{figure}

\subsection*{Multi-loop TDBS data handling}
For a given starting nominal liquid thickness, a time-averaged TDBS scan was first recorded. After completion of the scan, the liquid layer was thinned by displacing the piezo-driven stage by a 10~nm step towards the generation substrate, thereby reducing the liquid thickness by 10~nm. We assume that the change in liquid thickness equals the piezo displacement. After each displacement step, the system was left to stabilize and equilibrate for approximately 1--2~min before recording the next TDBS scan. This sequence, consisting of one TDBS acquisition followed by one 10~nm displacement step, was repeated until the glass lens came into direct mechanical contact with the chromium-coated substrate. The zero-thickness mechanical contact reference state was identified from the thickness dependence of the extracted Brillouin phase thanks to real time data treatment. For non-zero liquid thicknesses, the phase varies systematically from one thickness to the next, yielding a non-zero phase slope as a function of nominal thickness. Once liquid thickness reaches zero, further approach of the substrate no longer changes the phase. The disappearance of the phase slope was used as the criterion to define the direct mechanical contact point and thus the corresponding zero-thickness reference. Knowing the zero-thickness reference and the full displacement of the piezo stage from the initial thickness to contact, we can calibrate the absolute liquid thickness $d$ of the full sequence. The complete sequence from the initial thickness down to contact constitutes one experimental loop, yielding one full data set of Brillouin signals as a function of liquid thickness for a given sample. In our experiments, each loop typically covers a thickness range of 150--300~nm and was recorded in about half an hour. 

After completion of one loop, the piezo stage was retracted and the initial thickness for the next loop was set to $d_{\mathrm{initial}} - 1$~nm, where $d_{\mathrm{initial}}$ denotes the initial thickness of the previous loop. For example, if the first loop started at 210~nm, the second loop was initiated at 209~nm. This procedure was repeated nine times, yielding nine partially overlapping loops. Each loop sampled the thickness range with a nominal 10~nm step, while the 1~nm offset between successive loops provided finer effective sampling over the full data set. This multi-loop, coarse-step acquisition strategy was chosen instead of a single continuous fine-step thickness scan to reduce the influence of slow drifts, essentially caused by thermal fluctuations or mechanical vibrations. Lateral drift was assessed from lateral TDBS scans performed after the first and last loops. These measurements confirmed that the pump--probe overlap remained fixed relative to the contact region throughout the multi-loop acquisition.

An example of the multi-loop TDBS acquisition is shown in Fig.~\ref{sfigure2}. The figure presents the extracted phase of the 47.4~GHz Brillouin oscillations as a function of liquid thickness. Figure~\ref{sfigure2}A shows the phase data obtained from the nine acquisition loops. Within each loop, the liquid thickness was varied with a nominal piezo step of 10~nm, while the initial thickness was shifted by 1~nm between successive loops. The Brillouin phase was extracted from the time-domain signals by fast Fourier transform analysis around the 47.4~GHz frequency component. To obtain a higher-resolution thickness dependence, the data points from all nine loops were merged onto a common thickness axis, effectively interleaving the measurements from the different loops. Although each individual loop was acquired with a 10~nm nominal step, the 1~nm offset between successive loops provides a denser sampling of the thickness range. The resulting composite data set, shown in Fig.~\ref{sfigure2}B, therefore provides the Brillouin phase as a function of liquid thickness with 1~nm effective sampling. Note that a complete multi-loop acquisition required approximately 2~h. A finer effective thickness sampling can be achieved by reducing the displacement offset between successive loops and increasing the number of loops, albeit at the cost of longer acquisition times. For example, an effective sampling of 0.5~nm was obtained for 8CB, although such resolution was not necessary here because the confinement-induced variations occur over much longer length scales.

\begin{figure}[tb!]
\centering
\includegraphics[width=0.5\columnwidth]{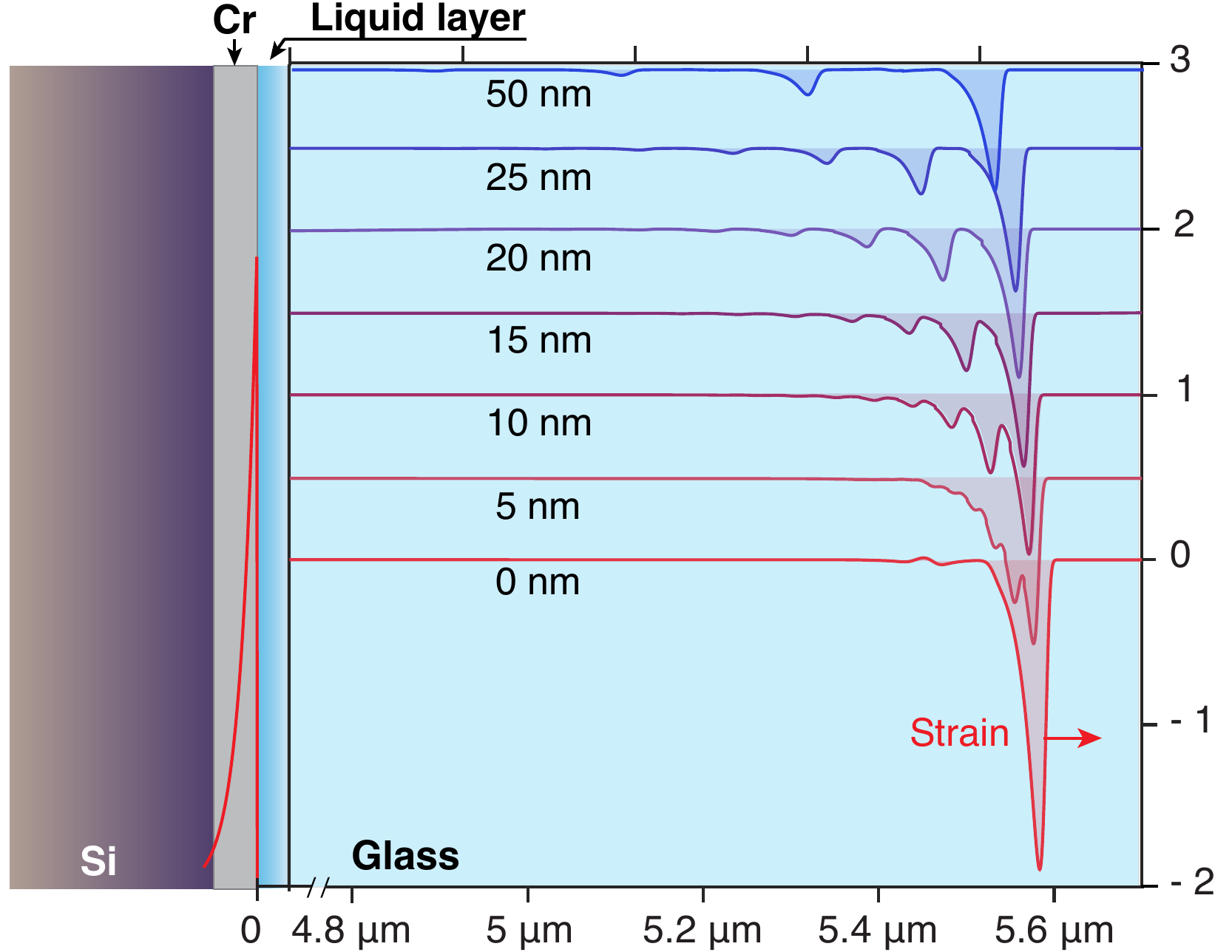}
\caption{Numerical simulation of the strain field transmitted into the glass lens. The strain profiles are shown at a delay of approximately 1~ns after laser excitation in the chromium transducer, for several liquid thicknesses. Increasing the liquid thickness produces a train of transmitted strain pulses resulting from multiple acoustic reverberations within the confined liquid layer.}
\label{simulations}
\end{figure}

\subsection*{Numerical modeling of acoustic propagation and optical detection}

Numerical simulations were performed to model the generation, propagation, and optical detection of acoustic pulses in the Cr/liquid/glass multilayer structure. Acoustic propagation was first calculated using one-dimensional time-domain simulations implemented with the k-Wave toolbox \cite{kwave}. The simulated structure consisted of the chromium transducer, the confined liquid layer, and the glass lens. For chromium, the density, longitudinal sound velocity, and refractive index were taken as 7150~kg~m$^{-3}$, 5940~m~s$^{-1}$, and 2.014-i$\times$3.46, respectively. For the glass lens, the corresponding values were 2196~kg~m$^{-3}$, 5968~m~s$^{-1}$, and 1.532. The spatial resolution was set to 0.2~nm and the temporal step to 5~fs.

Because the pump pulse is absorbed predominantly in the chromium film, acoustic excitation was modeled as an instantaneous pressure rise localized within the optical penetration depth of chromium. The generated strain pulse was then allowed to propagate through the multilayer structure, undergoing partial reflection and transmission at each acoustic interface. The resulting strain field was calculated as a function of time and position for each liquid thickness. Representative strain profiles transmitted into the glass lens, evaluated at a delay of approximately 1~ns for liquid thicknesses between 0 and 50~nm, are shown in Fig.~\ref{simulations}. As the liquid thickness increases, the transmitted acoustic field evolves into a train of pulses arising from multiple reverberations within the liquid cavity. When the delay between successive pulses becomes commensurate with the Brillouin period detected in the glass, the optical response is enhanced. This resonance is therefore acousto-optic in origin: it does not result from the buildup of overlapping acoustic echoes, but from enhanced Brillouin detection of temporally separated acoustic pulses.

Optical detection was modeled in a second step using the simulated acoustic strain fields as input. The transient reflectivity was calculated with a multilayer transfer-matrix method \cite{Orfanidis}. The sample was discretized into optical cells with the same spatial resolution as that used for the acoustic simulations. At each time delay, the strain-induced refractive-index modulation was included through the photoelastic relation $\Delta n = \left(\partial n/\partial \varepsilon_{zz}\right)\varepsilon_{zz}$, and the corresponding dynamic reflection coefficient was computed in the glass detection medium. This combined acoustic and optical model yields the calculated time-resolved reflectivity signal as a function of liquid thickness, allowing direct comparison with the experimental TDBS data.

The photoelastic coefficient of glass at the probe wavelength was taken as $\partial n/\partial \varepsilon_{zz} = -0.5$ \cite{Matsuda2002}. Because this coefficient is of order unity, the measured Brillouin amplitude provides an estimate of the strain transmitted into the glass. The extracted strain amplitudes lie in the range $10^{-4}$--$10^{-3}$, indicating that the acoustic excitation remains weak. The experiments and simulations therefore probe the linear acoustic response of the confined liquid, far from regimes in which strain-induced nonlinearities or irreversible structural modifications are expected \cite{Klieber2015}.

\section*{Acknowledgments}
We thank Keith Nelson from the Massachusetts Institute of Technology for his generous donation of a complete Coherent RegA laser system, without which this research would not have been possible. We thank Lionel Guilmeau for engineering support, Olivier Noel for advice on chromium deposition, and Mathieu Edely for performing AFM measurements of the surface roughness. We are grateful to Christoph Klieber for assistance with LabVIEW programming. R\'{e}mi Busselez and Vitaly Gusev are warmly acknowledged for their support during the initial stage of this work.

This work was funded by the Agence Nationale de la Recherche under grant ANR-22-CE42-0001 GigaSpin, by R\'{e}gion Bretagne through the SAD grant CHOCONDE, and by financial support from Rennes M\'{e}tropole.

\section*{Author Contributions}
TP, IC conceived the project and the experiments; the TDBS setup was build by IC, who conducted experiments, collected and analyzed the data. IC designed the samples. TP conducted numerical simulations. The manuscript was written by IC, TP with contribution from both authors.

\section*{Competing Interests}
No competing interests.

\section*{Correspondence}
\textsuperscript{1}To whom correspondence should be addressed. E-mail: ievgeniia.chaban@cnrs.fr, thomas.pezeril@cnrs.fr


\end{document}